\documentclass{emulateapj}

\usepackage{ifthen}
\newboolean{emulateapj}
\setboolean{emulateapj}{true}

\newcommand{\mbold}[1]{\mbox{\boldmath${#1}$}}

\usepackage{amsmath}

\ifthenelse{\boolean{emulateapj}}{
	\usepackage{epsf} 
	\usepackage{epsfig,floatflt}
	\usepackage{graphics}
}

\newcommand{\etal}{et~al.\ }
\newcommand{\etc}{etc.\ }
\newcommand{\ie}{i.e.,\ }
\newcommand{\eg}{e.g.,\ }

\begin{document}

\title{On the Viability of Bianchi Type VII$_h$ Models with Dark Energy}
\author{T. R. Jaffe\altaffilmark{1}, S.\ Hervik\altaffilmark{2}, A.\
J.\ Banday\altaffilmark{1}, K.\ M.\ G\'orski\altaffilmark{3} }  
\date{}

\altaffiltext{1}{Max-Planck-Institut f\"ur Astrophysik,
Karl-Schwarzschild-Str.\ 1, Postfach 1317, D-85741 Garching bei
M\"unchen, Germany; tjaffe@MPA-Garching.MPG.DE,
banday@MPA-Garching.MPG.DE.}

\altaffiltext{2}{Dept. of Mathematics and Statistics, Dalhousie
University, Halifax, Nova Scotia, B3H 3J5, Canada;
herviks@mathstat.dal.ca} 

\altaffiltext{3}{JPL, M/S 169/327, 4800 Oak Grove Drive, Pasadena CA
91109; Warsaw University Observatory, Aleje Ujazdowskie 4, 00-478
Warszawa, Poland; Krzysztof.M.Gorski@jpl.nasa.gov}

\begin{abstract}
We generalize the predictions for the CMB anisotropy patterns arising
in Bianchi type VII$_h$ universes to include a dark energy
component. We consider these models in light of the result of
\cite{jaffe:2005a,jaffe:2005b} in which a correlation was found on
large angular scales between the \emph{WMAP} data and the anisotropy
structure in a low density Bianchi universe.
We find that by including a term 
$\Omega_\Lambda > 0$, the same best-fit anisotropy
pattern is reproduced by several combinations of cosmological
parameters. This sub-set of models can then be
further constrained by current observations that limit
the values of various cosmological parameters. In particular,
we consider the so-called geometric degeneracy in these parameters
imposed by the peak structure of the \emph{WMAP} data itself.
Apparently, despite the additional 
freedom allowed by the dark energy component, 
the modified Bianchi models are ruled out at high significance.
\end{abstract} 

\keywords{
cosmology: cosmic microwave background -- cosmology: observations
}

\section{Introduction}

While cosmology appears to be converging on a ``concordance model''
that describes the universe as inflationary and isotropic, there
remain unexplained anomalies in the CMB data, and other models are not
yet ruled out.  The \emph{WMAP} data provide some of the most accurate
measurements yet of the cosmic microwave background and contribute
to high accuracy determinations of cosmological parameters
\citep{bennett:2003,spergel:2003}. However, there are several studies
that show that at large scales, the CMB is not statistically
isotropic and Gaussian, as predicted by inflation theory \citep{de
Oliveira-Costa:2004, eriksen:2004, hansen:2004b, vielva:2004}.

In \cite{jaffe:2005a,jaffe:2005b}, we examined a particular set of
anisotropic cosmological models: the Bianchi type VII$_h$ class of
spatially homogeneous generalizations of Friedmann universes that
include small vorticity (universal rotation) and shear (differential
expansion) components.  Surprisingly, we found evidence that one of
these models correlates with the CMB sky and that subtracting this
component resolves several of these observed anomalies.  The models
used in that study were derived by \cite{BJS} and include no
cosmological constant component in the total energy density.  The
best-fit model found by that study required
$\Omega_{\textrm{m}0}=0.5$, implying, $\Omega_k=0.5$, \ie a
significantly negatively curved universe.

\citet{land:2005} subsequently considered flat Bianchi models 
and sought explicitly to 
resolve the problems of the low quadrupole and the low-$\ell$
alignments using a statistic constructed to achieve that purpose.
Their analysis does indeed find a model
that fixes these anomalies, 
but it remains statistically insignificant as a detection.
Our original result has the benefit of being a detection that is based
entirely on a simple least-squares fit of the Bianchi template
to the data, yet we find that it serendipitously
resolves several anomalies 
without that requirement having been built-in to the search algorithm.  
The importance of our 
result, therefore, lies in 
the fact that it resolves the problems of the low quadrupole,
low-$\ell$ alignments, power asymmetry, and non-Gaussian cold spot.

However, both analyses neglect the fact that 
the existing Bianchi solutions include no dark energy component.
Furthermore, the best-fit results require an energy content that is
consistent neither with other astronomical observations 
nor with the CMB itself.  
In this work, we present a modification of the Barrow \etal Bianchi
type VII$_h$ solution so that it includes a cosmological constant term
in the evolution.  We discuss the impact of that term on the structure
of the resulting CMB anisotropy pattern, particularly the degeneracy
that is introduced in the model space by the addition of
$\Omega_\Lambda$.  We then discuss the viability of $\Lambda$ models
that are morphologically identical to our best-fit $\Omega_\Lambda=0$
models by considering the constraints imposed by different
measurements of the cosmological parameters. 


\section{Bianchi Models with $\Lambda > 0$}\label{sec:models}


\subsection{Solution}

First we shall generalize the equations in \cite{BJS} to include a
cosmological constant $\Lambda$. The basic assumption is that the type
VII$_h$ universe is close to FRW.

We start with the equations of motion using expansion-normalized
variables \citep[see, \eg ][]{DynSys}. The equations describing the
evolution of Bianchi type VII$_h$ universes with a tilting perfect
fluid can be found in \cite{CH} and \cite{HHC}.  In our case, we
consider a tilting perfect fluid with $\gamma=1$ (dust) and a
cosmological constant. The equations in \cite{CH} can easily be
generalized to include a cosmological constant by adding a matter term
$(\Omega_{\Lambda},\gamma_{\Lambda}=0)$ wherever there is a matter
term $(\Omega,\gamma)$.

Furthermore, we assume that the universe is close to a FRW at all
times and that the tilt velocity of the dust fluid is small; hence we
assume:
\begin{eqnarray}
\Sigma^2\ll 1, ~|{\bf N}_{\times}|\ll 1, ~ v_1\ll 1, ~|{\bf v}|\ll 1.
\end{eqnarray}
Here, $\Sigma^2=\sigma^2/(3H^2)$ is the expansion-normalized shear
scalar, ${\bf N}_{\times}$ is a complex curvature variable, and $v_1$,
${\bf v}=v_2+iv_3$ are the tilt components (see \citealt{CH} for
details).  The curvature variables $A$ and $\bar{N}$ need not be
small, and they are related to the group parameter $h$ as follows:
\[ A^2\approx 3h\bar{N}^2. \] 
We also assume that the parameter $h$ is not too small: $h\geq
\mathcal{O}(1)$.  (As $h\rightarrow 0$, the assumptions made for this
derivation break down and a qualitatively different solution for type
VII$_0$ is needed.  See \citealt{HHLC}.)

Given the above assumptions, the deceleration parameter is 
\begin{eqnarray}
q\approx \frac 12\Omega_{\textrm{m}}-\Omega_{\Lambda},
\end{eqnarray}
and the equation of motion for the Hubble scalar is
\begin{eqnarray}
H'=-(q+1)H,
\end{eqnarray}
where the prime indicates the derivative with respect to the
dimensionless time $T$ defined by ${dt}/{dT}= 1/H$, and $t$ is the
cosmological time.

At the lowest order, the tilt components that are related to the
vorticity are the components $v_2+iv_3\equiv{\bf v}$ \citep[in the
notation of ][]{CH}. These induce non-zero shear components
$\Sigma_{12}+i\Sigma_{13}\equiv{\mbold \Sigma}_1$ via the linear
constraint equation, which in the FRW limit reduces to:
\begin{eqnarray}
{\mbold \Sigma}_1\bar{N}(i-3\sqrt{h})+\Omega_{\textrm{m}}{\bf v}=0.
\label{constraint}
\end{eqnarray}
Moreover, close to FRW the curvature $K\equiv A^2$ and matter
densities $\Omega_{\textrm{m}}$ and $\Omega_{\Lambda}$ are related via the
Friedmann equation
\begin{eqnarray}
1=\Omega_{\Lambda}+K+\Omega_{\textrm{m}}.
\end{eqnarray}
We also define $x$ 
to be
\begin{eqnarray}
x\equiv\left(\frac{h}{K_0}\right)^{\frac 12}=\left(\frac{h}{1-\Omega_{\Lambda 0}-\Omega_{\textrm{m0}} }\right)^{\frac 12}.
\end{eqnarray}
Eq. (\ref{constraint}) now reduces to Barrow et al.'s eqs.(4.6) and
(4.7) by dropping the subscript $0$ and replacing $\Omega$ with
$\Omega_{\textrm{m}}$. The equations of motion can now be solved.

By choosing $T_0=0$, $x=e^{\alpha_0}H_0$, we can relate $z$ to $T$
and the time variable $\tau$ in \cite{BJS} to the redshift $z$:
\begin{eqnarray}
1+z&=&e^{-T}, \\
\frac{d T}{d \tau}&=& He^{T}e^{\alpha_0}. 
\end{eqnarray}

The Hubble scalar, $K$, $\Omega_{\textrm{m}}$, and $\Omega_{\Lambda}$ can then be written
\begin{equation}
H(z)=H_0\mathcal{H}(z),
\end{equation}
\begin{equation}
\mathcal{H}(z)= \left[\Omega_{\Lambda 0}+K_0(1+z)^2+\Omega_{\textrm{m}0}(1+z)^3\right]^{\frac 12},\label{eq:h_of_z}
\end{equation}
\begin{equation}
K = \frac{K_0(1+z)^2}{\mathcal{H}^2(z)}, 
\label{eq:sigbjorn12}
\end{equation}
\begin{equation}
 \Omega_{\textrm{m}} = \frac{\Omega_{\textrm{m}0}(1+z)^3}{\mathcal{H}^2(z)},  
\end{equation}
\begin{equation}
 \Omega_\Lambda = \frac{\Omega_{\Lambda0}}{\mathcal{H}^2(z)}, \label{eq:lambda}
\end{equation} 
where $\Omega_{\Lambda0},K_0,\Omega_{\textrm{m}0}$ are the present values of
the expansion-normalized $\Lambda$, curvature and ordinary matter
(dust) and obey the constraint:
\[ 1=\Omega_{\Lambda0}+K_0+\Omega_{\textrm{m}0}.\]
 We also note that $K=\Omega_k>0$ implies negative curvature. The
 shear components $\Sigma_{12}$ and $\Sigma_{13}$, and the remaining
 curvature variable are given by
\begin{equation}
\Sigma_{12}=\frac{\Sigma_{12,0}(1+z)^3}{\mathcal{H}(z)}, \label{eq:Sigma}
\end{equation}
\begin{equation}
 \Sigma_{13}=\frac{\Sigma_{13,0}(1+z)^3}{\mathcal{H}(z)},
\end{equation}
\begin{equation}
\bar{N} = \frac{\bar{N}_0(1+z)}{\mathcal{H}(z)}.
\end{equation}


Conformal time becomes
\begin{equation}
\tau - \tau_0 = -\frac{1}{x}\int_0^z \frac{dz}{\mathcal{H}(z)}, 
\end{equation}
which can be used in the Barrow expressions for $s$ and $\psi$.  
The integrals of Barrow \etal equations (4.12) and (4.13) become
\begin{equation}
\int_0^{z_E} \frac{ s(1-s^2)\sin{\psi}(1+z)^2dz}{(1+s^2)^2\mathcal{H}(z)}
\end{equation}
\begin{equation}
\int_0^{z_E} \frac{ s(1-s^2)\cos{\psi}(1+z)^2dz}{(1+s^2)^2\mathcal{H}(z)}
\end{equation}
with $C_3=4$, where $z_E$ is the redshift at photon emission.  In
calculating the constant $C_1$, the density $\Omega_0$ should now be
$\Omega_{\textrm{m}0}$, i.e. the matter density only.

The following assumptions should be kept in mind regarding this
solution:
\begin{itemize}
\item The tilt velocity is small.  In a dust- or $\Lambda$-dominated
Bianchi type VII$_h$ universe, the tilt asymptotically tends to zero at
late times.  Using the above relations, it can also be shown that for
the case of interest, it remains small back to the last scattering
surface as well.

\item The universe is close to FRW at all times.  This assumption is
required to make the problem tractable.  In a $\Lambda$-dominated
universe, the type VII$_h$ models become close to FRW.  For the cases
in question, plugging the numbers into the above relations shows that
this assumption still holds at the surface of last scattering.  (At
very early times, however, this is not the case.)  It is likely that
models for which this assumption does not hold at last scattering
would then have greater shear effects, but the details would require a
more complicated derivation.

\item \cite{BJS} only consider the contribution from the
vorticity that directly involves the shear variables
$\Sigma_{12}$ and $\Sigma_{13}$. The (expansion-normalized) shear
scalar, $\Sigma^2=\sigma^2/(3H^2)$, is given by
\begin{eqnarray}
\Sigma^2=\Sigma_+^2+\left|{\mbold\Sigma}_{\times}\right|^2+\Sigma_{12}^2+\Sigma_{13}^2
\end{eqnarray}  
and involves therefore all of the shear components. In \cite{BJS} and
here, the other shear components have simply been ignored in the
calculation of $\Delta T/T$. 
Considering additional shear degrees of freedom is in principle
possible but in practice difficult.  The effect of such additions is
one of the largest unknowns in this analysis.

\end{itemize}

\subsection{Properties}\label{sec:properties}

In the presence of the cosmological constant, a universe that is very
nearly flat now and that was very nearly flat at the time of last
scattering can become significantly negatively curved at intermediate
redshifts.  We can see this from Eqtn. \ref{eq:sigbjorn12} above.  

Figure \ref{fig:k_z} shows how $K(z)$ evolves for different values of
$\Omega_{\Lambda 0}$ and $\Omega_{\textrm{m}0}$.  At high redshift, the matter
dominates and the curvature vanishes, and at low redshift,
$\Omega_\Lambda$ can dominate, depending on its exact value.  At
intermediate redshifts, however, the universe may become more
negatively curved.  For a currently almost flat universe with
parameters close to the observed values of $\Omega_{\Lambda 0}\sim
0.7$ and $\Omega_{\textrm{m}0}\sim 0.3$, the curvature is at most only a few
percent (black curve).  Only for very small matter densities and very
large dark energy densities does the curvature become large at
intermediate redshifts.  To reproduce the observed asymmetry with
these $\Lambda>0$ models requires either a large current negative
curvature (pink and blue dot-dashed curves) or a very small matter
density (green dashed curve).

\ifthenelse{\boolean{emulateapj}}{
\begin{figure}[h]
{\epsfxsize=\linewidth \epsfbox{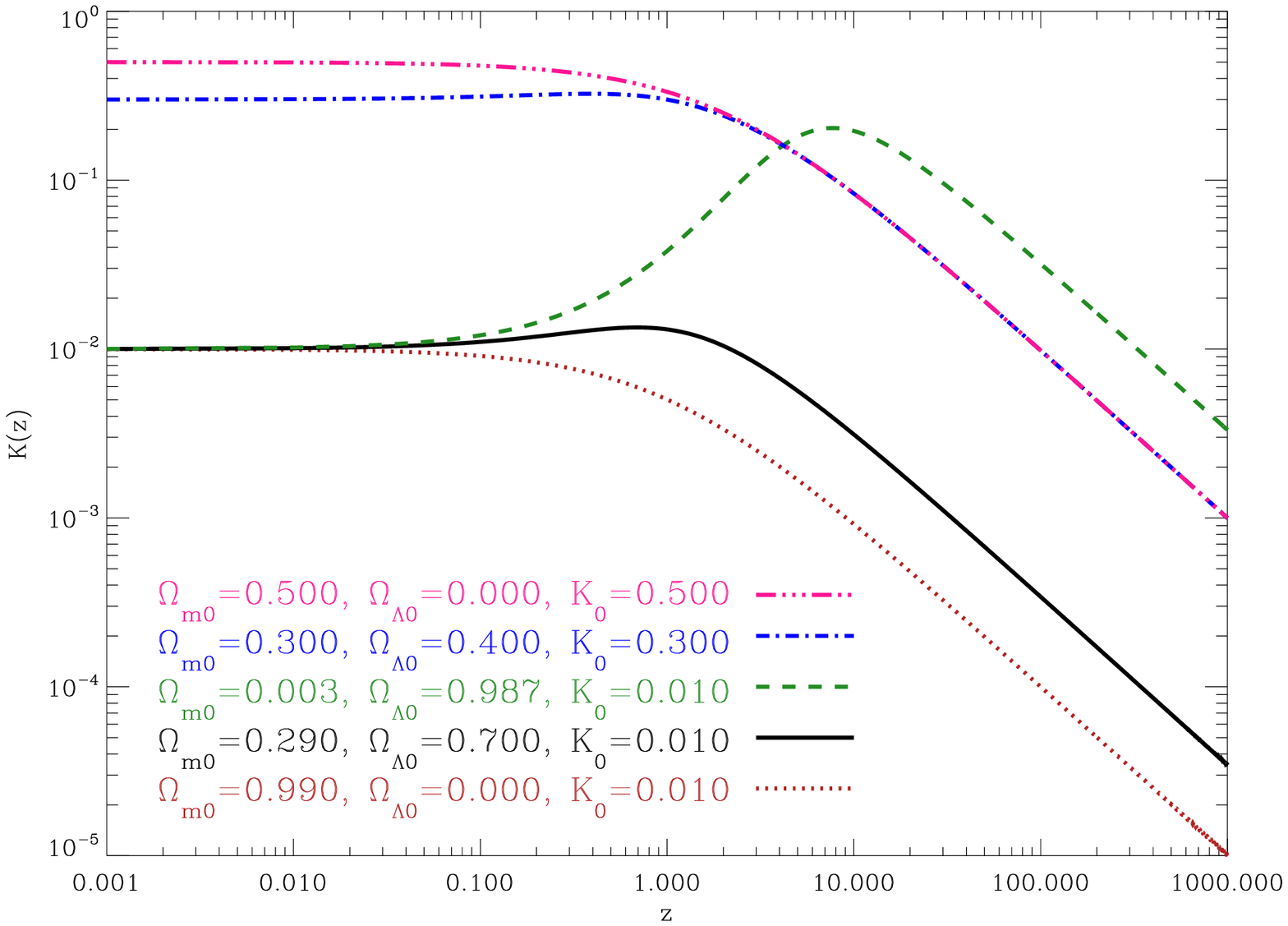}}
\caption{ Examples of the evolution of $K(z)$.  See text.}
\label{fig:k_z}
\end{figure}
}

The addition of a $\Lambda$ component adds a third parameter to the
Bianchi parameter space, but there is a degeneracy to the resulting
morphology of the Bianchi-induced pattern.  The same structure in the
induced anisotropy pattern can be reproduced by different parameter
combinations.  The addition of dark energy results in a tighter spiral
for a given value of $x$, because the redshift to the surface of last
scattering corresponds to a larger time difference.  The geodesics
have therefore completed more rotations since recombination for a
given value of $x$ when $\Omega_\Lambda > 0$.  Secondly, the focusing
of the spiral depends on the curvature between the observer and the
last scattering surface.  Significant focusing can arise from a
universe that is very open now, or from a universe where $\Lambda$ is
large enough to give negative curvature at intermediate redshifts (as
shown in Fig. \ref{fig:k_z}.)

Three combinations of parameters, for example, can reproduce the same
apparent structure on the sky :
\begin{enumerate}
\item \label{item:one}
$(x,\Omega_{\Lambda0},\Omega_{\textrm{m0}})=(0.62,0,0.5)$ -- \ie no 
$\Lambda$ and a matter content half critical, giving a large current
negative curvature \citep[the model found in ][]{jaffe:2005b};
\item \label{item:two}
$(x,\Omega_{\Lambda0},\Omega_{\textrm{m0}})=(0.8,0.4,0.3)$ -- \ie approximately
the observed matter content (baryon plus dark) , with some $\Lambda$
but still a large current negative curvature;
\item  \label{item:three}
$(x,\Omega_{\Lambda0},\Omega_{\textrm{m0}})=(4.0,0.987,0.003)$ -- \ie a much
smaller than observed matter content, but a nearly flat current
curvature ($\Omega_k=0.01$); the large $\Lambda$ causes a large
negative curvature at intermediate redshifts.
\end{enumerate}
The real difference between these models is the amplitude:
(\ref{item:three}) has an amplitude $\sim80\%$ of (\ref{item:one}),
and (\ref{item:two}) has an amplitude of $\sim 8$ {\it times} that of
(1).  The shear and vorticity values obtained from fitting these
templates to the data would change accordingly.

Effectively, this implies that a given structure on the sky
characterized by the amount of geodesic focusing and the number of
spiral turns is the same for all models along a line in the three
dimensional parameter space of $(\Omega_{\textrm{k0}},
\Omega_{\Lambda0}, x)$.  The degeneracy in the templates is
only broken by the amplitude of the variation.
But $(\sigma/H)_0$ is what we measure by fitting a template to the
sky, so we cannot distinguish among the degenerate models without an
independent measurement of the shear.

It is not straightforward to calculate where such lines of degeneracy
lie.  Instead, we determine this empirically by generating a grid of
models and comparing them to the previously determined best-fit model.

\section{Viability of a Bianchi Type VII$_h$ Model}\label{sec:viable}

We now address the question of whether the degeneracy of the Bianchi
type VII$h$ models with $\Omega_{\Lambda 0}>0$ described
in \S \ref{sec:models} gives us enough freedom to define a Bianchi
model that is morphologically identical to the best-fit model in
\citet{jaffe:2005a,jaffe:2005b} and that is also consistent with estimates
of the cosmological parameters.

\subsection{Matter Density and 'Geometric' Degeneracy}\label{sec:geo_degen}

\cite{efstathiou:1999} describe the limitations of constraining
cosmological parameters with CMB data alone.  In particular, there is
a 'geometric' degeneracy in the curvature and dark energy density (or
equivalently, the matter density.)  Degenerate models have: 1) the
same values of the physical baryon density , $\omega_{\textrm{b}}=\Omega_{\textrm{b}} h^2$,
and cold dark matter density, $\omega_{\textrm{c}}=\Omega_{\textrm{c}} h^2$; 2) the same
primordial fluctuation spectrum; and 3) the same value of the
parameter
\begin{equation}
\mathcal{R}=\sqrt{\frac{\omega_{\textrm{m}}}{\omega_k}}
\begin{cases}
\sinh[\sqrt{\omega_k}y] & \text{if }\omega_k > 0, \\
\sqrt{\omega_k}y        & \text{if }\omega_k = 0, \\
\sin[\sqrt{\omega_k}y]  & \text{if }\omega_k < 0,
\end{cases}
\end{equation}
where
\begin{equation}
y=\int_{a_{\textrm{rec}}}^{1} \frac{da}{\sqrt{\omega_{\textrm{m}} a + \omega_k a^2 +
\omega_\Lambda a^4}}
\end{equation}
(where $a_{\textrm{rec}}$ is the scale factor at recombination).  These
models will have almost the same power spectra.  Contours of constant
$\mathcal{R}$ in model space $\Omega_k$ versus $\Omega_\Lambda$ are
shown in Figure \ref{fig:degen}, with red highlighting the curve
intersecting the best-fit, flat, \emph{WMAP}-only parameters
\citep[Table~1]{spergel:2003}. Figure
\ref{fig:degen_cls} shows power spectra for those degenerate models
(calculated using CMBFAST.\footnote{{\tt http://www.cmbfast.org/}})
When normalized, the spectra are identical at the acoustic peaks, and
the degeneracy only fails at large angular scales.

\ifthenelse{\boolean{emulateapj}}{
\begin{figure}
{\epsfxsize=\linewidth \epsfbox{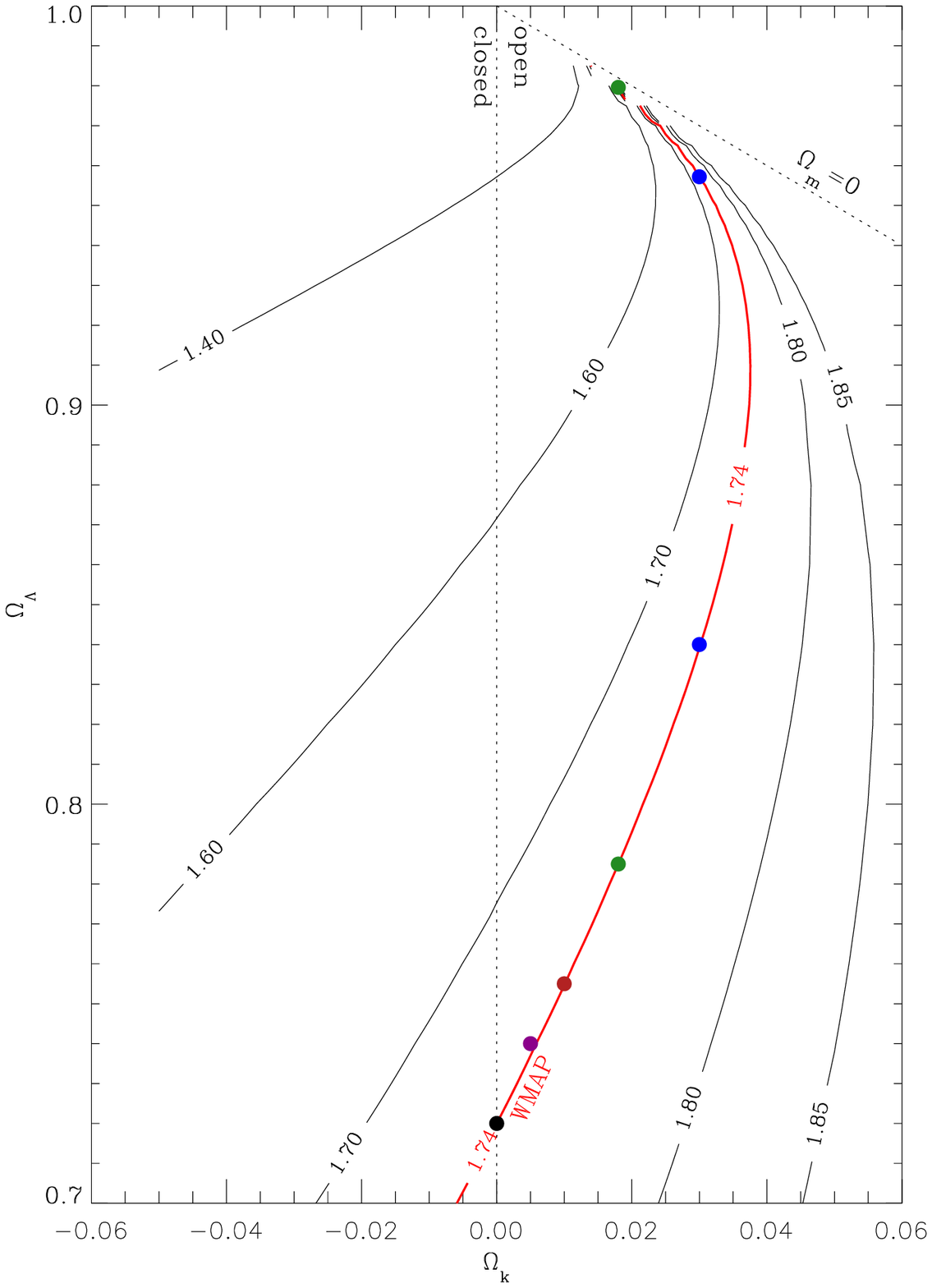}}
\caption{ Degeneracy contours on the $\Omega_\Lambda-\Omega_k$
plane for $\omega_{\textrm{b}}=0.023$ and $\omega_{\textrm{m}}=0.13$.  Along the red line
denoting $\mathcal{R}$ for the \emph{WMAP} best-fit parameters, the
circles mark example power spectra that are shown in Figure
\ref{fig:degen_cls}.}
\label{fig:degen}
\end{figure}
}

\ifthenelse{\boolean{emulateapj}}{
\begin{figure}
{\epsfxsize=\linewidth \epsfbox{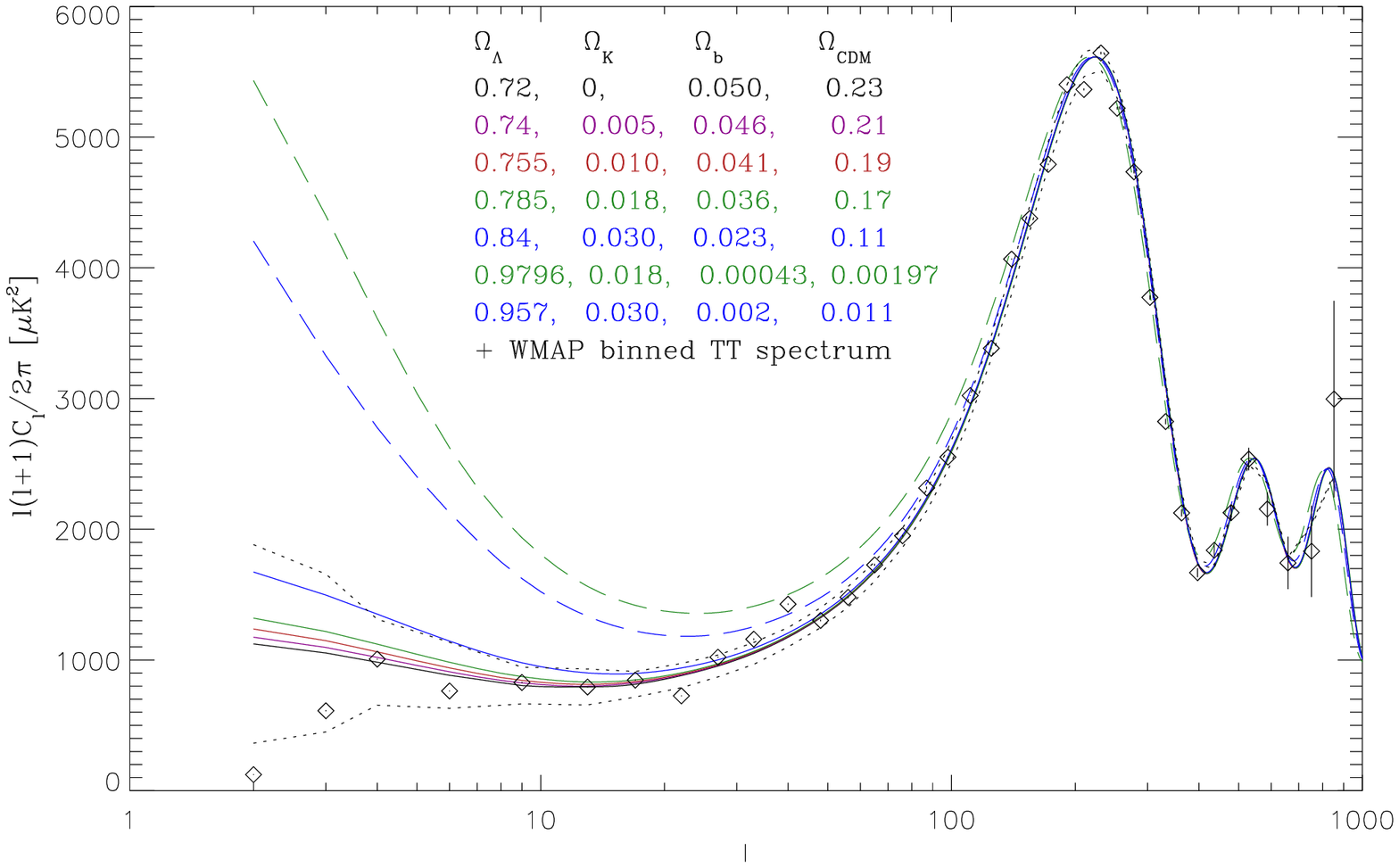}}
\caption{ Power spectra for several models degenerate with the
\emph{WMAP} data (marked with circles in figure \ref{fig:degen}.)  The
solid versus dashed green and blue curves are where there 
are two degenerate models at the same value of $\Omega_k$; the dashed
is that with the higher $\Omega_\Lambda$.  The spectra are normalized
so that the amplitude at the first acoustic peak is always the same.
The dotted black line shows the cosmic variance uncertainty for each
data point around the \emph{WMAP} model.}
\label{fig:degen_cls}
\end{figure}
}

In the three dimensional parameter space of $(\Omega_{k0},
\Omega_{\Lambda 0}, x)$, the geometric degeneracy forms a surface,
while the Bianchi models with identical structure form a line.  The
two need not intersect, but in the case of the observed \emph{WMAP}
CMB-only power spectrum of \cite{spergel:2003} and best-fit Bianchi
model of \cite{jaffe:2005a,jaffe:2005b}, they do.  There is a Bianchi
model that has the identical structure to the best-fit
$\Omega_\Lambda=0$ model and lies on the
\emph{WMAP} degeneracy curve.  It has parameters
$(\Omega_{k0},\Omega_{\Lambda0}, x)=(0.028,0.96,2.5)$.  But
how viable is this region of parameter space?

As shown in Figure \ref{fig:degen_cls}, the geometric degeneracy is
broken at large angular scales.  Models with high $\Omega_{\Lambda0}$
and low $\Omega_{\textrm{m0}}$ predict too much large scale
power. \emph{WMAP} data alone place relatively loose constraints on
$\Omega_{\textrm{m0}}$, but even these rule out such low values as
required for the Bianchi models.  (The power spectrum of the sky
corrected for the Bianchi component has less large-scale power overall
\citep{jaffe:2005a,jaffe:2005b} than the uncorrected power shown in
that figure, so the problem will only become worse.)

To quantify these limits, we examine the posterior likelihood from the
\emph{WMAP} data alone using the COSMOMC\footnote{{\tt
http://cosmologist.info/cosmomc/}} code of \cite{lewis:2003}, which
implements a Monte Carlo Markov Chain method.  For a simple look to
compare the
\emph{WMAP} constraints to the Bianchi degeneracy, we explore the
five-parameter space of: $\theta$ (the ratio of the sound horizon to the
angular diameter distance), $\tau$ (optical depth), $\Omega_k$
(curvature), $n_{\textrm{s}}$ (spectral index of scalar perturbations),
and $A_{\textrm{s}}$ (amplitude of scalar perturbations), with fixed parameters
$\omega_{\textrm{b}}=0.023$ (the physical baryon density),
$\omega_{\textrm{dm}}=0.107$ (the physical dark matter density),
$f_\nu=0$ (neutrino fraction of dark matter density), 
$w=-1$ (dark energy equation of state), $r\equiv
A_{\textrm{t}}/A_{\textrm{s}}=0$ (ratio of tensor to scalar fluctuations).
Other parameters such as $H_0$, $\Omega_{\Lambda0}$, and
$\Omega_{\textrm{m0}}$ are derived from this set.  (The free parameters
have only such priors as defined by ranges that are much broader than
any realistic uncertainty.)  The resulting constraints in the
$\Omega_{\textrm{m}}-\Omega_{\Lambda}$ plane (marginalizing over the other
parameters) are shown in Figure \ref{fig:om_ol_space} in red.  All
Bianchi models with the structure of the best-fit model lie well
outside the 95\% confidence region from the \emph{WMAP} data alone.
Allowing the physical matter densities to vary as well but including
other datasets, the constraints (in magenta) are even tighter.

\cite{tegmark:2004} give $\Omega_{\textrm{m0}}=0.57^{+0.45}_{-0.33}$
from \emph{WMAP} data alone, which tightens to
$\Omega_{\textrm{m0}}=0.40^{+0.10}_{-0.09}$ adding data from the Sloan
Digital Sky Survey (SDSS).
See also \cite{sanchez:2005} for CMB+2dFGRS results.

\ifthenelse{\boolean{emulateapj}}{
\begin{figure}[t]
{\epsfxsize=\linewidth \epsfbox{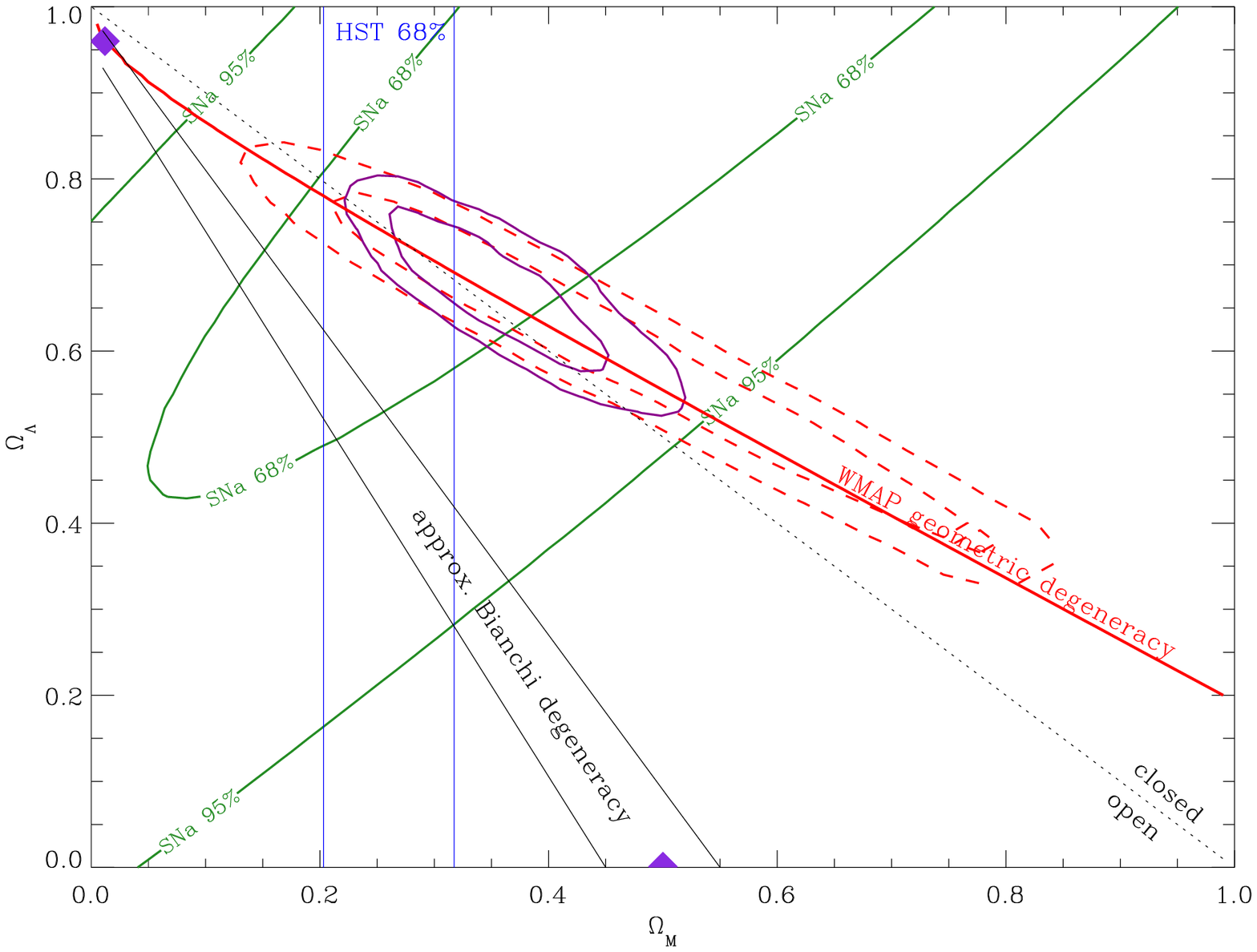}}
\caption{ Parameter space by $\Omega_{\textrm{m}}$.  In green are
the supernovae constraints \citep{knop:2003}.  The solid red line is
the same geometric degeneracy curve shown in Figure
\ref{fig:degen}. In blue are the HST Key Project \citep{freedman:2001}
constraints from $\Omega_{\textrm{m0}}=\omega_{\textrm{m}}/h^2$ (assuming
fixed $\omega_{\textrm{m}}$ and $\omega_{\textrm{c}}$).  The black solid lines show an
approximate representation of where the Bianchi degeneracy lies, from
the original model at $(x,\Omega_{\Lambda0},\Omega_{\textrm{m}0})=(0.62,0,0.5)$
to the one that lies on the \emph{WMAP} geometric degeneracy curve at
$(0.028,0.96,2.5)$ (each shown with a violet diamond.)  The likelihood
contours from
\emph{WMAP} data alone (computed using cosmomc; see text) are shown
with the red dashed lines.  In solid magenta are the contours from
\emph{WMAP}, supernovae, HST, and SDSS data combined, where $\omega_{\textrm{m}}$
and $\omega_{\textrm{c}}$ also vary.}
\label{fig:om_ol_space}
\end{figure}
}

\subsection{Optical Depth}

Figure \ref{fig:degen_cls} shows that the geometric degeneracy breaks
at large angular scales, and the models with high $\Omega_\Lambda$
have too much large scale power.  But there is an additional
degeneracy if we allow the optical depth, $\tau$, to vary.  Adjusting
$\tau$ and $A_{\textrm{s}}$ such that $A_{\textrm{p}}\equiv
A_{\textrm{s}}e^{-2\tau}$ remains constant has exactly the effect we
need of modifying only the large scale power while leaving the peak
heights unchanged
\citep{tegmark:2004}.  Figure \ref{fig:cls_tau} shows how this works,
but that the effect is not large enough.  Even a $\tau$ of zero does
not bring the power at large angular scales within range of the data,
and that model is inconsistent with the large-scale peak in the TE
spectrum.

One can also ask if the addition of the Bianchi component can affect
the variance of the large scale TE cross-power, resulting in an
incorrect estimate of $\tau$ due to a chance alignment of the
polarization signal with the Bianchi structure.  Simulations with no
reionization and with an added Bianchi temperature pattern (the same
location and shear amplitude as our best-fit model) show that this is
not the case.  The cross-power remains flat at low $\ell$ with the
expected variance. 

Note that \cite{hansen:2004a} find different values of $\tau$ derived
from fitting temperature data in the Northern and Southern hemispheres
as defined in the frame of reference that maximizes the power
asymmetry (see \citealt{hansen:2004b}.)  The additional large-scale
structure in the South that the Bianchi template reproduces could
cause this effect.  If a polarization signal were also produced in a
Bianchi geometry and correlated with the temperature anisotropy, it
would influence the measured optical depth from the low-$\ell$ TE peak
as well.

\ifthenelse{\boolean{emulateapj}}{
\begin{figure}
{\epsfxsize=\linewidth \epsfbox{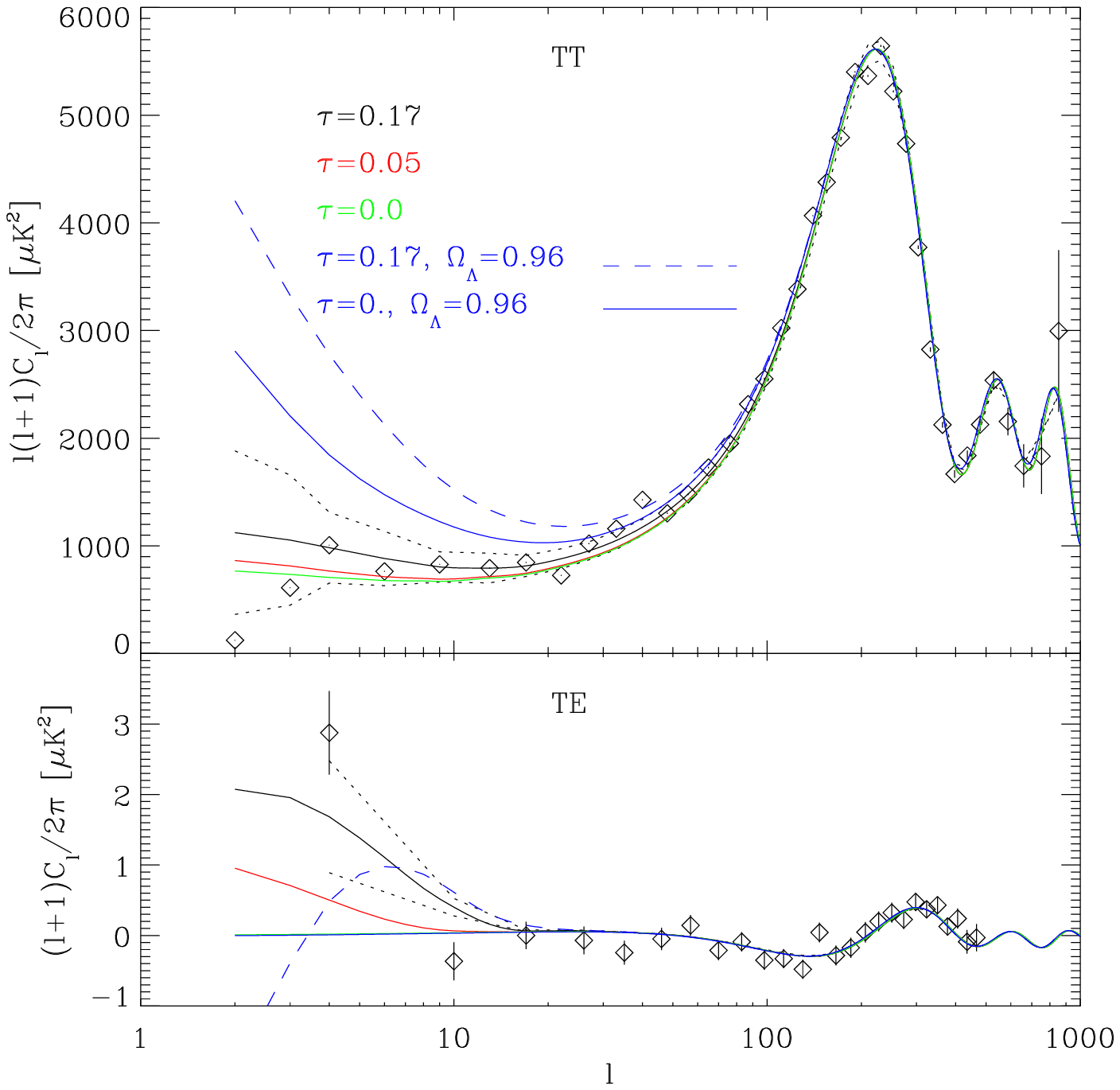}}
\caption{ Power spectra as a function of optical depth, $\tau$, each
normalized such that the first peak heights are all the same
(effectively changing $A_s$.)  The top panel is the TT power, and the
bottom the TE cross-power.  Lowering $\tau$ lowers the large scale
power, but not enough to bring the high $\Omega_\Lambda$ models down
to the range of the data. } 
\label{fig:cls_tau}
\end{figure}
}

\subsection{Supernovae, $H_0$, \etc}

The degeneracy contours in Figure \ref{fig:degen} are plotted for
constant values of $\omega_{\textrm{m}}=0.13$ and $\omega_{\textrm{b}}=0.023$ \citep[from
][]{spergel:2003}.  For different points in the
$(\Omega_{k0},\Omega_{\Lambda 0})$ space, $\Omega_{\textrm{m}0}$
changes and therefore the value of $h\equiv H_0/(100\ \textrm{km
s}^{-1} \textrm{Mpc}^{-1})$ changes, increasing with increasing
$\Omega_{\Lambda 0}$ along the degeneracy curve.  For models of high
$\Omega_{\Lambda 0}$ that lie on the geometric degeneracy curve, $H_0$
reaches values over $300$.  This is ruled out at high significance by
the \emph{WMAP} data itself; \citet{tegmark:2004} give
$H_0=48^{+27}_{-12} \textrm{km s}^{-1}\textrm{Mpc}^{-1}$.  Independent
determinations of Hubble's constant also rule out these values, \eg
the HST Key Project \citep{freedman:2001} value of $72\pm 8 \textrm{km
s}^{-1}\textrm{Mpc}^{-1}$.  See, e.g., \citet[Table~3]{spergel:2003}
for other estimates of the Hubble constant.

High-redshift supernovae observations can break the geometric
degeneracy by independently placing constraints that are nearly
perpendicular to the \emph{WMAP} constraints in
$\Omega_{\textrm{m}}-\Omega_\Lambda$ space (see Figure
\ref{fig:om_ol_space}.) The degeneracy curve for the best-fit Bianchi
model compared to the \emph{WMAP} data lies outside the $98\%$
confidence contour determined by the Supernova Cosmology Project
\citep[fit~no.~6]{knop:2003}.

It is worth pointing out that it is not clear whether type Ia
supernovae are truly standard candles out to high redshift.  The
uncertainty, however, is not enough to accommodate such low values of
$\Omega_{\textrm{m0}}$.

%

The addition of cosmic vorticity and shear would, of course, influence
the determinations of such parameters.  The sky coverage of supernova,
for example, is fairly extensive outside the Galactic Plane region,
but the sample size is small.  Studies such as
\citet{kolatt:2001} have ruled out significant dipole or
quadrupole asymmetry in the expansion, but a test for a more
complicated anisotropy structure induced by the vorticity would
require far more supernova in the sample and a good sky coverage.
Furthermore, as the current value of the shear expansion is very
small, distance measurements to relatively low red-shift (compared to
the CMB) objects may not be sensitive enough to detect it.  This also
implies that such a small current shear would not significantly alter
measurements of other cosmological parameters such as the Hubble
constant that are dependent on relatively low-redshift observations.

\subsection{Other Bianchi Models}

One can also ask, then, how much one can vary the parameters of the
Bianchi model and still have a statistically significant fit.  The
most interesting region on Figure \ref{fig:om_ol_space} is near
$(\Omega_{\textrm{m}},\Omega_\Lambda)=(0.15,0.75)$, where the
\emph{WMAP} and supernovae constraints approach the region of the
best-fit Bianchi model.  This model resembles the best-fit model, but
since it is closer to flat than the Bianchi degeneracy curve, there is
less geodesic focusing.  As a result, the model is not so good a fit
to the data, and has a significance, compared to Gaussian
realizations, of only $85\%$, compared to over $99\%$ for the model at
the same $\Omega_\Lambda$ but on the Bianchi degeneracy curve.

%

Considering that the \emph{WMAP} data somewhat favor a closed
universe, one might ask about closed models with vorticity and shear.
These are Bianchi type IX models, also discussed in \citet{BJS}.
Unlike the open type VII$_h$ models, however, closed models exhibit
neither geodesic focusing nor the spiral pattern, even in the presence
of vorticity.  Barrow \etal use them to place limits on vorticity and shear
simply using the quadrupole.  Such closed models do not reproduce the
morphology needed to explain the power asymmetry or the cold spot, and
it would be impossible to claim any detection with only the quadrupole
as an observable.

\subsection{Other Dark Energy Models}

The cosmological constant is the simplest form of dark energy, and all
observations so far remain consistent with it.  Other models are not
ruled out, however, and in some cases are favored.  Alternatives come
in many varieties, some with physical motivation, others constructed
to give a particular result.  (See, \eg \citealt{padmanabhan:2003}.)
Here, we address the question of
whether an alternative dark energy model can bring our Bianchi pattern
any closer to the constraints imposed by the data. 

Dark energy models are characterized by their equation of state,
$p=w\rho$, where a cosmological constant $\Lambda$ corresponds to a
model with a constant $w=-1$.  Alternative theories allow $w$ to vary
with time, as in ``quintessence'' and ``k-essence'' models.
A rather {\it ad hoc} parameterization is often used of the form
$w=w_0+w_1z$, which allows comparison of generic dark energy models
with supernova data \citep{wang:2004,dicus:2004}.

The derivation in \S \ref{sec:models} can be generalized for any dark
energy model by the appropriate substitution into equations
\ref{eq:h_of_z} and \ref{eq:lambda} of $\Omega_X$ instead of
$\Omega_\Lambda$.  We then have
\begin{equation}
\mathcal{H}(z)^2=\Omega_{X0}f_X(z) + K_0(1+z)^2 + \Omega_{\textrm{m}0}(1+z)^3 
\end{equation}
where the function $f_X$ is derived from the equation of state as
\begin{equation}
f_X=\textrm{exp}\left\{ 3\int_0^z \frac{dz'}{1+z'} \left(1+w(z')\right)\right\}
\end{equation}

Examples for extreme values of $w_0$ and $w_1$ are shown in Figure
\ref{fig:k_z_de}.  As was shown in Figure \ref{fig:k_z}, we can look
at the evolution of 
the curvature as a function of time for these models and see if any
display the required negative curvature at intermediate redshift.
Using this parameterization, or alternatively the parameterization
$w=w_0+w_1\frac{z}{z+1}$, does not give much additional freedom,
however, to create a viable model.  Ultimately, the amount of
curvature and therefore focusing is still largely a function of the
relative densities of matter and dark energy and of the current
curvature.  Manipulating the evolution of the dark energy equation of
state simply changes the time scales on which the transitions between
the different components of the density occur.

\ifthenelse{\boolean{emulateapj}}{
\begin{figure}[h]
{\epsfxsize=\linewidth \epsfbox{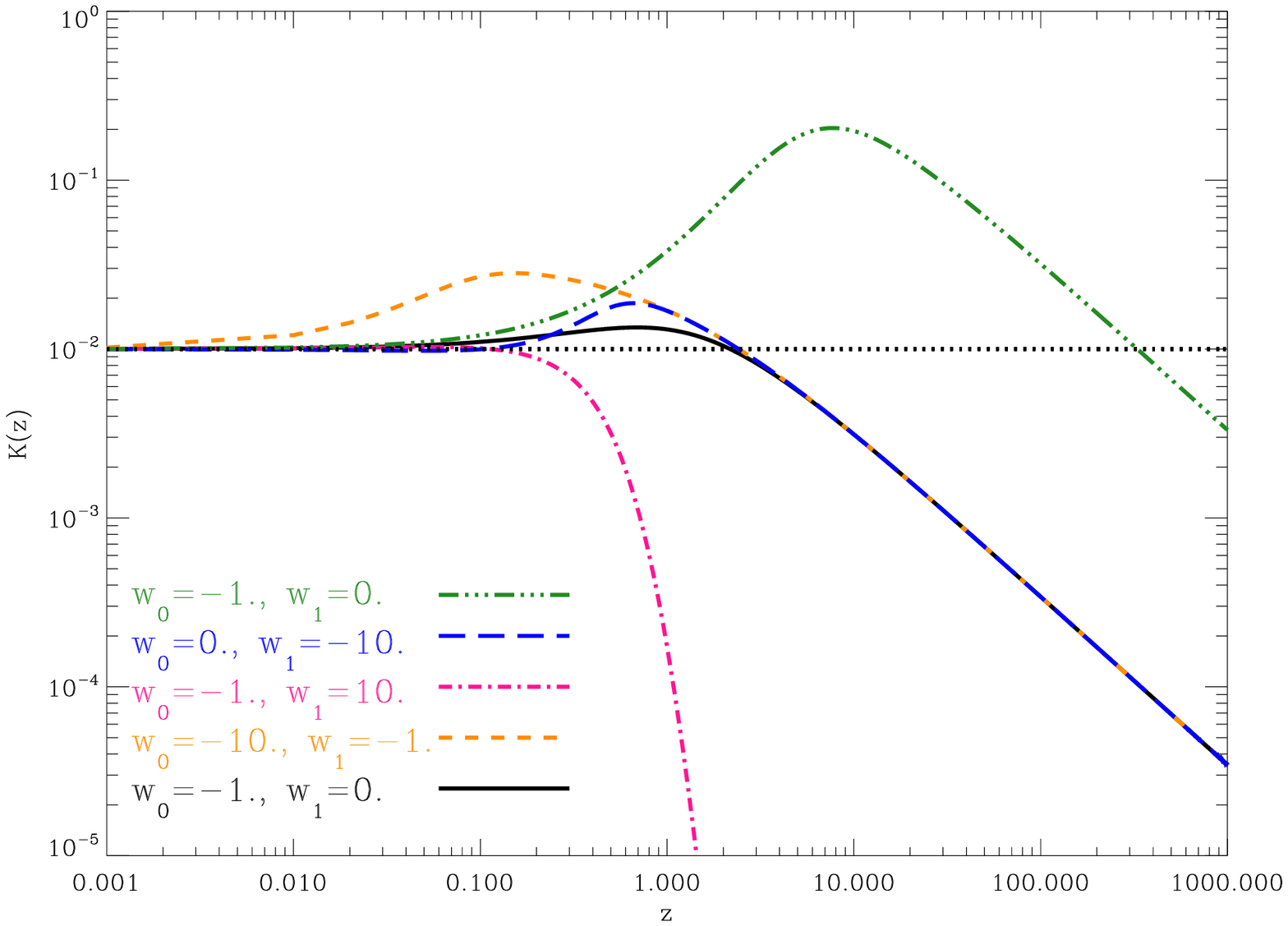}}
\caption{ Examples of the evolution of $K(z)$ for different dark
energy models where the current curvature is constrained to be small
($K_0=0.01)$.  $\Omega_{\textrm m0}=0.29$ for all except the green
triple-dot-dashed curve, where $\Omega_{\textrm m0}=0.003$.  This is
the curve shown in Fig. \ref{fig:k_z} with a small matter density,
high $\Omega_\Lambda$, and approximately the amount of curvature
needed to create the best-fit Bianchi model.}
\label{fig:k_z_de}
\end{figure}
}

\subsection{Small-scale Structure}

We have so far ignored an additional issue with these models related
to the stochastic component of the CMB.  The analysis in
\citet{jaffe:2005a,jaffe:2005b} assumes that the observed CMB anisotropies
consist of two independent signals: the predicted Bianchi pattern, and
a stochastic, statistically isotropic component.  The latter may be
generated via inflation or another mechanism, but if it is
statistically isotropic at the surface of last scattering, it might no
longer be statistically isotropic when observed after traveling
through a Bianchi universe.  


The power spectrum at small scales might deviate from predictions
due to additional structure (dependent on orientation relative to the
preferred axis) introduced by propagation in a Bianchi metric.  We
note that the asymmetry observed in \citet{eriksen:2004} and
\citet{hansen:2004b} extends to $\ell\sim 40$, while the Bianchi model
has structure only up to $\ell\sim 15$.  Geodesic focusing might cause
such a power asymmetry, though it might also require that the
asymmetry continue to the smallest angular scales, which is not
observed.  \citet{hansen:2004b} also find that some unexplained
outliers in the \emph{WMAP} power spectrum are associated with
different hemispheres.

Furthermore, the generation of the fluctuations at the last scattering
surface might also be affected by the anisotropy, though in the case
in question, that anisotropy remains very small ($\la 10^{-4}$) at
$z\sim 1000$.  Essentially, we expect that the Bianchi models could
also be constrained by the lack of deviations in the power spectrum
from the best-fit \emph{WMAP}/concordance model, although detailed
predictions for such deviations are needed.

The consistency of the acoustic peak scale with the flat concordance
model would be very difficult for our best-fit Bianchi model to
explain.

\subsection{Early Universe}

Considering the success of Big Bang Nucleosynthesis in explaining
light element abundances, any model that significantly changes the
physical processes at those early times must be ruled out.  Several
studies have thus attempted to place limits on shear expansion and
rotation by examining their effect on the relative Helium abundance,
$Y$.  \cite{barrow:1976} originally showed that such analysis can
place strong constraints on shear.  The interesting exception,
however, was type VII$_h$ models, where Barrow found that the CMB
remained a stronger constraint.  Later studies showed that including
more complicated effects can reverse the trend of $Y$ with the shear
(see, \eg \citealt{jbg:1983}).  \cite{barrow:1984} revisited the
issue and showed that in some cases, even extremely anisotropic models
may still have the observed Helium abundance.  Bianchi type VII$_h$
models, however, have not been treated in detail since the result of
\cite{barrow:1976}.  After demonstrating for several cases, not
including VII$_h$, that $Y$ increases for anisotropic models,
\cite{matzner:1986} {\it conjecture} that the same trend applies for
{\it all} anisotropic models, but they admit that this is difficult to
prove.

In short, it is possible that such Bianchi models are strongly ruled
out by BBN and the observed Helium fraction.  But this has not been
definitively proved.

\section{Discussion and Conclusions}

We have presented solutions for Bianchi type VII$_h$ type universes
that include a dark energy term and examined how their morphological
properties change over the parameter space.  The addition of dark
energy adds a degeneracy such that different combinations of the three
parameters $(\Omega_{\textrm{m}},\Omega_\Lambda, x)$ can lead to the
same observed structure as in the best-fit model of
\cite{jaffe:2005a,jaffe:2005b}.  A template can be constructed that
has the identical structure of that best-fit model and also falls on
the geometric degeneracy curve for the parameters as measured by
\emph{WMAP} data.

This model, however, lies well outside the 95\% confidence region in
$\Omega_{\textrm{m}}-\Omega_\Lambda$ space for \emph{WMAP} data, ruled
out by the over-prediction of large scale power.  It also lies outside
the 95\% likelihood contours from the Supernova Cosmology Project, and
is further inconsistent with the measurement of the Hubble constant
from the Hubble Key Project.  Bianchi models that are more consistent
with these other measurements are no longer good fits to the
\emph{WMAP} large scale structure.  

One of the most difficult problems for these models is to account for
the acoustic peak structure.  The anisotropy at early times might
influence the nature of the fluctuations at last scattering, and the
geometry could affect the power spectrum on small angular scales due
to the geodesic focusing between last scattering and the observer.
Detailed predictions for these effects would be needed, but it is
difficult to envision such an anisotropic scenario that happened to
reproduce the observed acoustic peak structure, mimicking the
concordance cosmology so well.

There is currently no prediction for the CMB polarization anisotropy
in a Bianchi universe, but such a geometry-induced signal could
provide an additional test of these models.  If the preferred direction
indicated in the temperature data is also reflected in the full sky
polarization data (expected from further \emph{WMAP} data releases
and, eventually, from Planck), there will be even more motivation to
consider non-standard models.

We have shown that our best-fit Bianchi type VII$_h$ model is not
compatible with measured cosmological parameters, despite the
additional freedom from adding dark energy.  It is worth reiterating,
however, that the serendipitous discovery of a theoretically derived
template that correlates well with the data also happens to resolve
several anomalies that cannot be explained in the standard picture.
These particular models may not be viable, but lacking any plausible
scenario for systematics or foregrounds to be the source of the
anomalies, non-standard models that reproduce a similar morphology
merit continued interest.



\section*{Acknowledgments}

We are grateful to S.~D.~M.~White for useful discussions and
suggestions.  SH was funded by a Killam PostDoctoral Fellowship.  We
acknowledge use of the HEALPix software \citep{healpix} and analysis
package for deriving some results in this paper.  We also acknowledge
use of the Legacy Archive for Microwave Background Data Analysis
(LAMBDA).

\ifthenelse{\boolean{emulateapj}}{}{
\clearpage
}

\ifthenelse{\boolean{emulateapj}}{
\end{document}
}

%
%

\begin{figure}[h]
\plotone{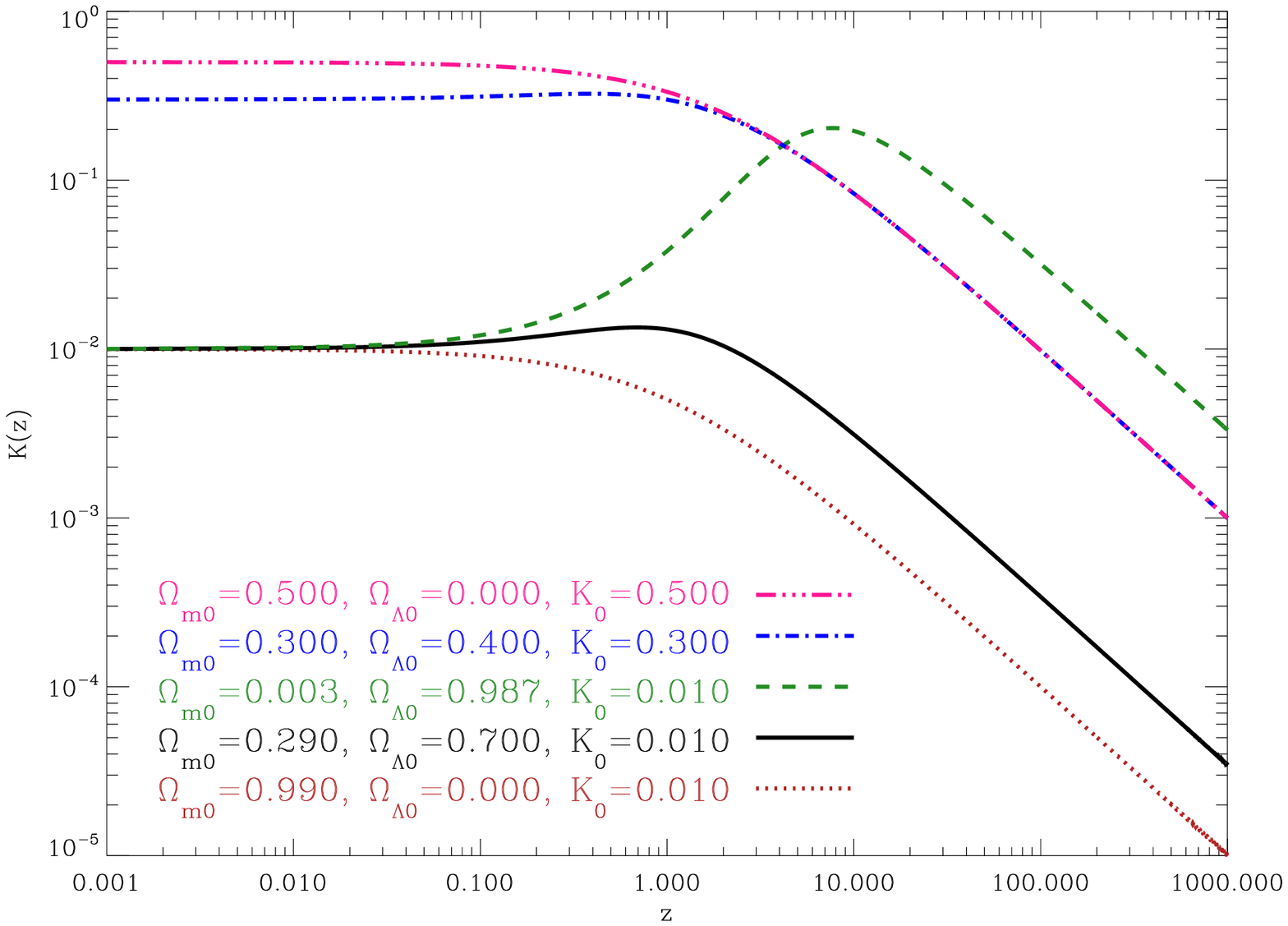}
\figcaption{ Examples of the evolution of $K(z)$.  See text.  \label{fig:k_z}}
\end{figure}

\begin{figure}
\plotone{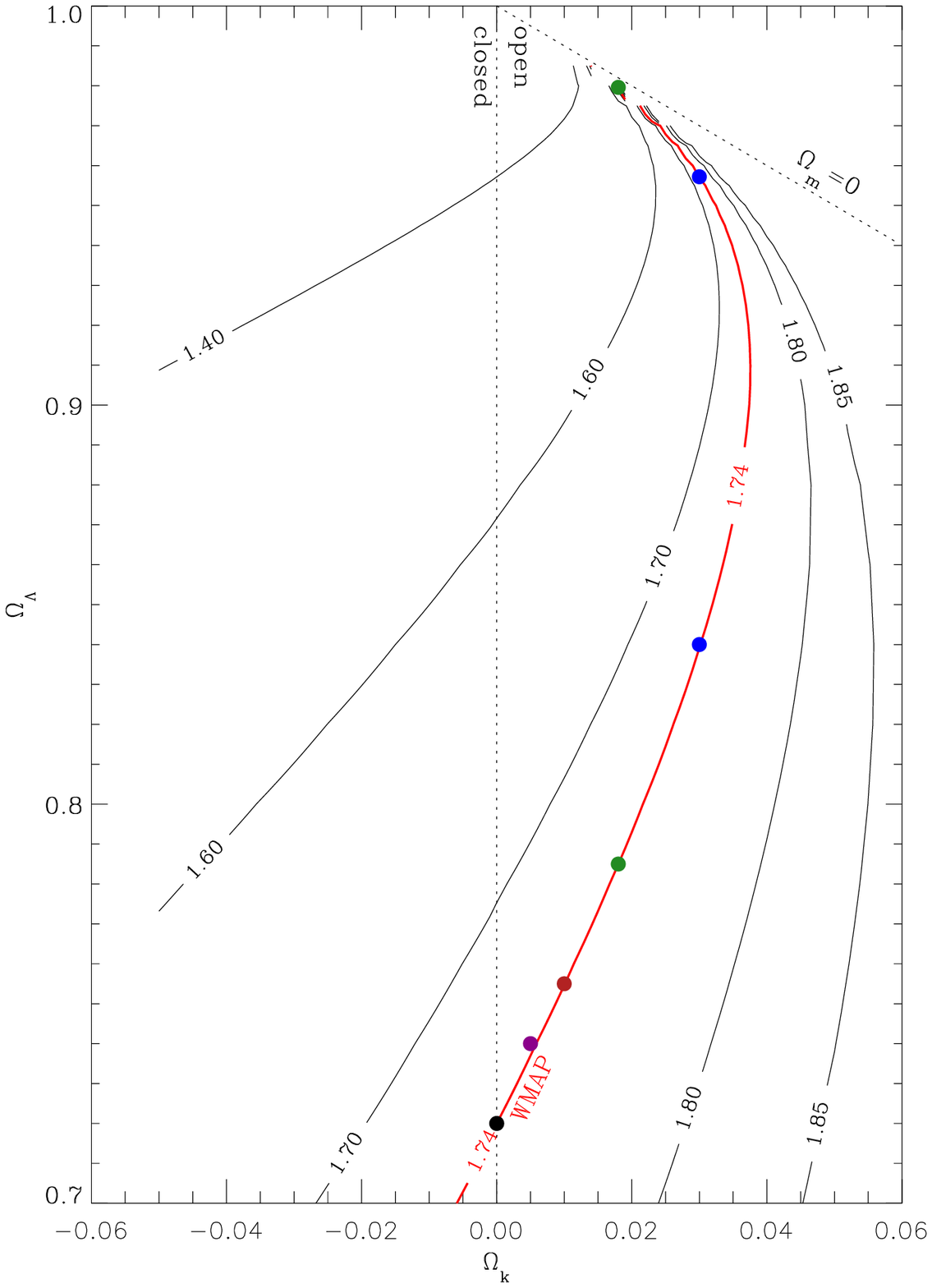}
\figcaption{ Degeneracy contours on the $\Omega_\Lambda-\Omega_k$
plane for $\omega_{\textrm{b}}=0.023$ and $\omega_{\textrm{m}}=0.13$.  Along the red line
denoting $\mathcal{R}$ for the \emph{WMAP} best-fit parameters, the
circles mark example power spectra that are shown in Figure
\ref{fig:degen_cls}.\label{fig:degen}}
\end{figure}

\begin{figure}
\plotone{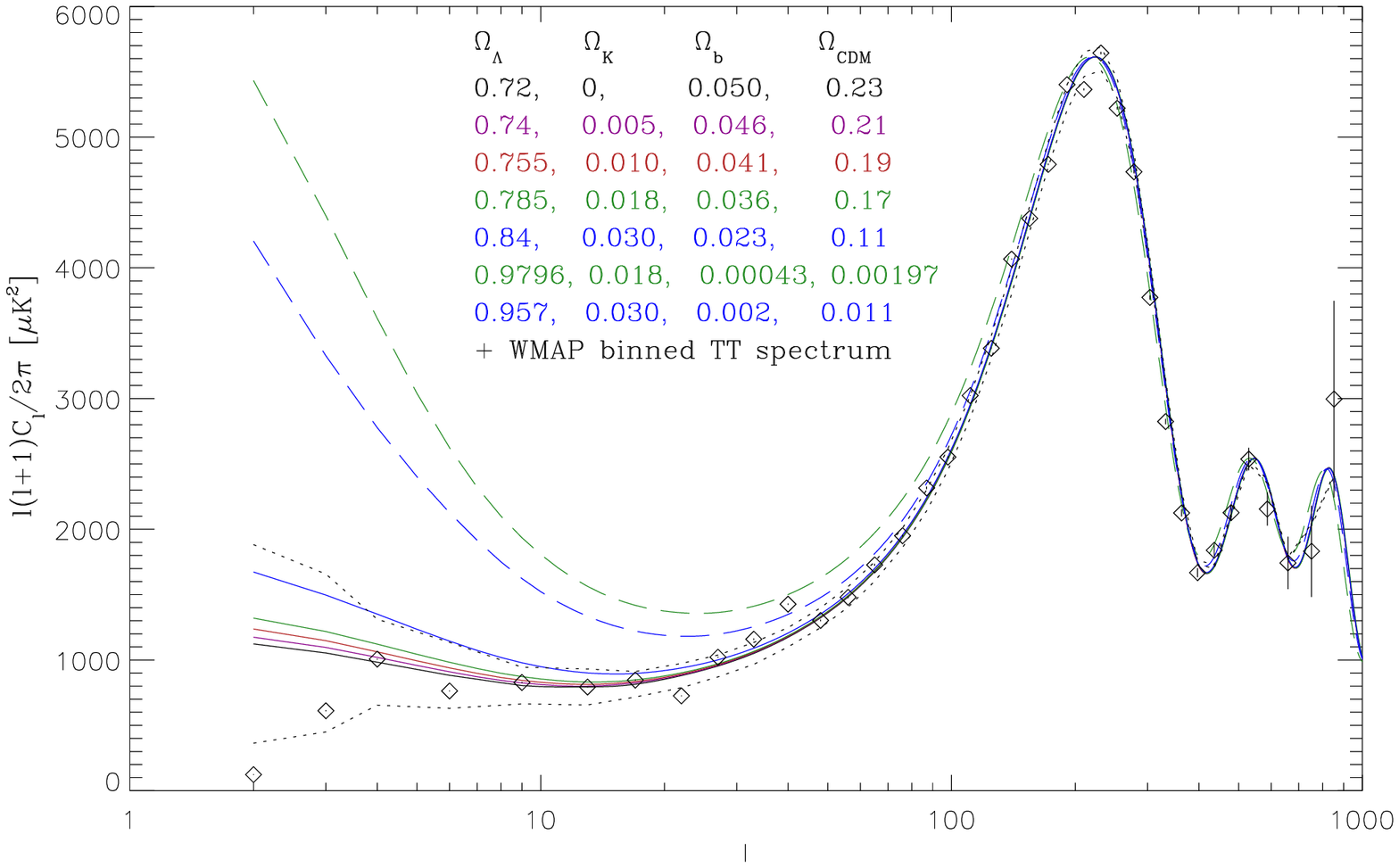}
\figcaption{ Power spectra for several models degenerate with the
\emph{WMAP} data (marked with circles in figure \ref{fig:degen}.)  The
solid versus dashed green and blue curves are where there 
are two degenerate models at the same value of $\Omega_k$; the dashed
is that with the higher $\Omega_\Lambda$.  The spectra are normalized
so that the amplitude at the first acoustic peak is always the same.
The dotted black line shows the cosmic variance uncertainty for each
data point around the \emph{WMAP} model.\label{fig:degen_cls}}
\end{figure}

\begin{figure}[t]
\plotone{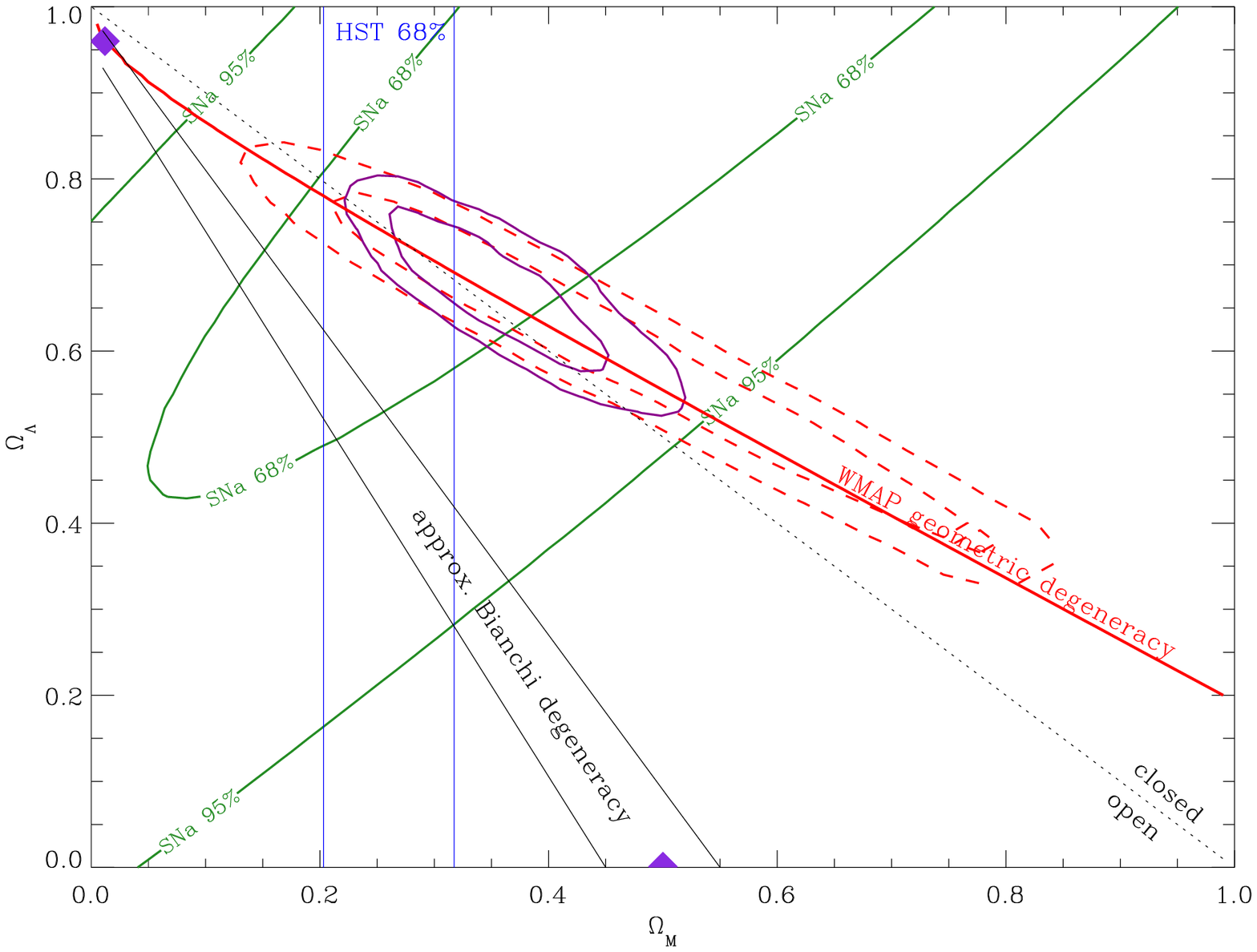}
\figcaption{ Parameter space by $\Omega_{\textrm{m}}$.  In green are
the supernovae constraints \citep{knop:2003}.  The solid red line is
the same geometric degeneracy curve shown in Figure
\ref{fig:degen}. In blue are the HST Key Project \citep{freedman:2001}
constraints from $\Omega_{\textrm{m0}}=\omega_{\textrm{m}}/h^2$ (assuming
fixed $\omega_{\textrm{m}}$ and $\omega_{\textrm{c}}$).  The black solid lines show an
approximate representation of where the Bianchi degeneracy lies, from
the original model at $(x,\Omega_{\Lambda0},\Omega_{\textrm{m}0})=(0.62,0,0.5)$
to the one that lies on the \emph{WMAP} geometric degeneracy curve at
$(0.028,0.96,2.5)$ (each shown with a violet diamond.)  The likelihood
contours from
\emph{WMAP} data alone (computed using cosmomc; see text) are shown
with the red dashed lines.  In solid magenta are the contours from
\emph{WMAP}, supernovae, HST, and SDSS data combined, where $\omega_{\textrm{m}}$
and $\omega_{\textrm{c}}$ also vary.\label{fig:om_ol_space}}
\end{figure}

\begin{figure}
\plotone{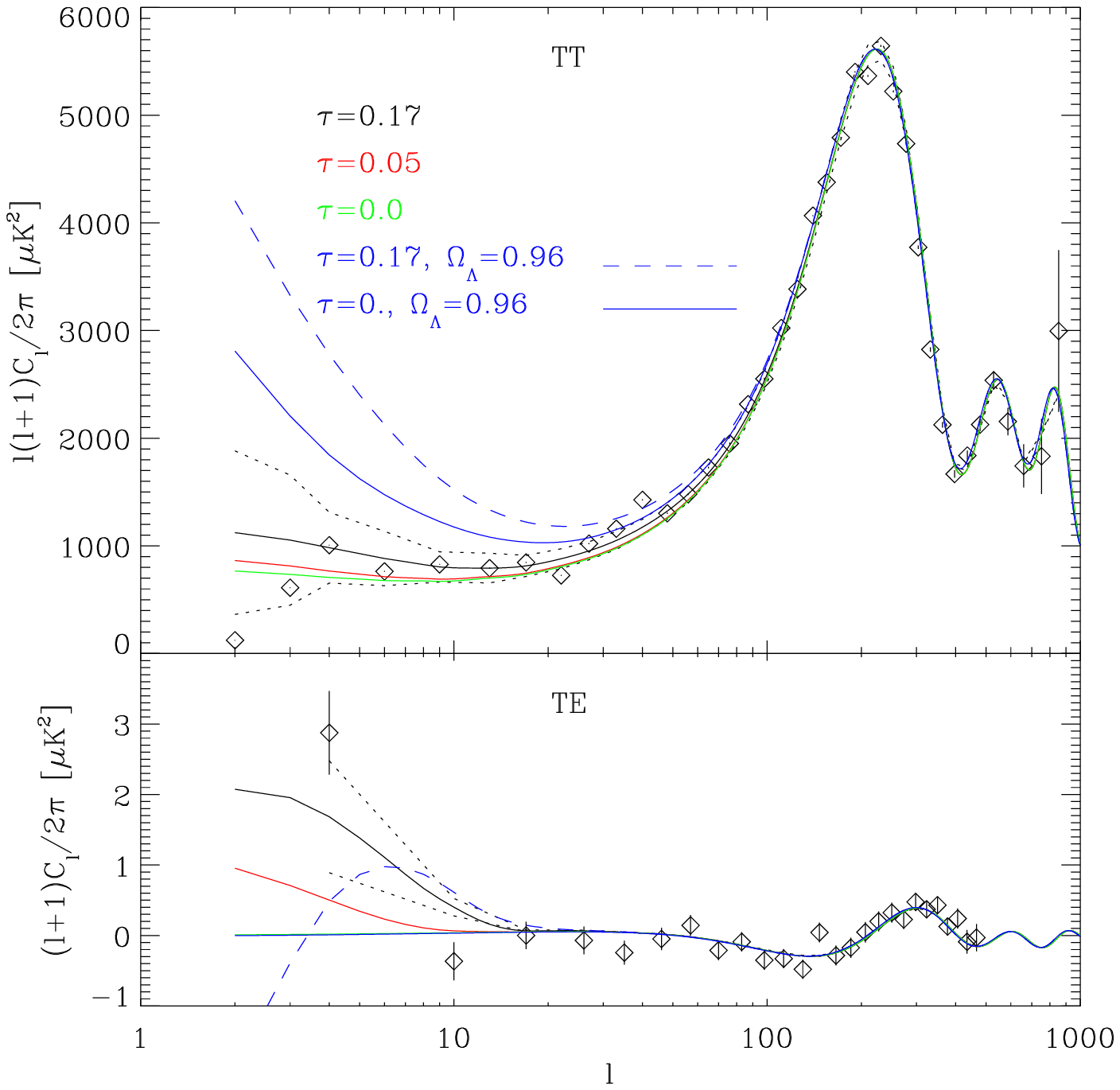}
\figcaption{ Power spectra as a function of optical depth, $\tau$, each
normalized such that the first peak heights are all the same
(effectively changing $A_{\textrm{s}}$.)  The top panel is the TT
power, and the bottom the TE cross-power.  Lowering $\tau$ lowers the
large scale power, but not enough to bring the high $\Omega_\Lambda$
models down to the range of the data. \label{fig:cls_tau}}
\end{figure}

\begin{figure}[h]
\plotone{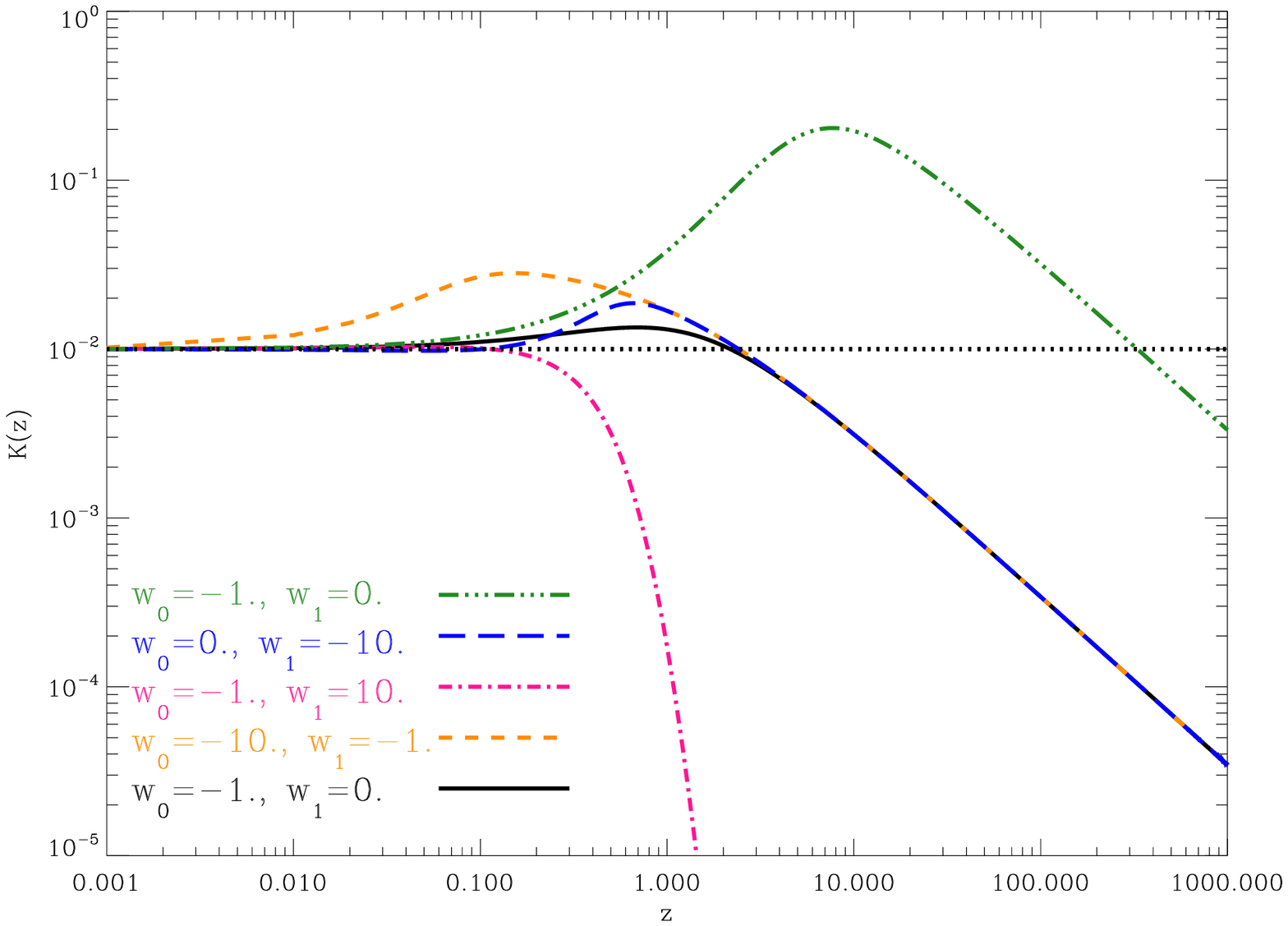}
\figcaption{ Examples of the evolution of $K(z)$ for different dark
energy models where the current curvature is constrained to be small
($K_0=0.01)$.  $\Omega_{\textrm m0}=0.29$ for all except the green
triple-dot-dashed curve, where $\Omega_{\textrm m0}=0.003$.  This is
the curve shown in Fig. \ref{fig:k_z} with a small matter density,
high $\Omega_\Lambda$, and approximately the amount of curvature
needed to create the best-fit Bianchi model. \label{fig:k_z_de}}
\end{figure}

\end{document}